\documentclass[aps,floats]{revtex4}
\usepackage{amsmath,amssymb}
\usepackage{graphicx,epsfig}

\begin{document}
\bibliographystyle {plain}

\def\oppropto{\mathop{\propto}} 
\def\opsimeq{\mathop{\simeq}}
\def\opoverderline{\mathop{\overline}}
\def\operarrow{\mathop{\longrightarrow}}
\def\opsim{\mathop{\sim}}

\def\fig#1#2{\includegraphics[height=#1]{#2}}
\def\figx#1#2{\includegraphics[width=#1]{#2}}


\title{  Random Transverse Field Ising model in $d=2$  :  \\
analysis via Boundary Strong Disorder Renormalization  } 


 \author{ C\'ecile Monthus and Thomas Garel }
  \affiliation{Institut de Physique Th\'{e}orique, 
CNRS and CEA Saclay, 
 91191 Gif-sur-Yvette cedex, France}

\begin{abstract}
To avoid the complicated topology of surviving clusters induced by standard Strong Disorder RG in dimension $d>1$, we introduce a modified procedure called 'Boundary Strong Disorder RG' where the order of decimations is chosen a priori. We apply numerically this modified procedure to the Random Transverse Field Ising model in dimension $d=2$. We find that the location of the critical point, the activated exponent $\psi \simeq 0.5$ of the Infinite Disorder scaling, and the finite-size correlation exponent $\nu_{FS} \simeq 1.3$ are compatible with the values obtained previously by standard Strong Disorder RG. Our conclusion is thus that Strong Disorder RG is very robust with respect to changes in the order of decimations. In addition, we analyze in more details the RG flows within the two phases to show explicitly the presence of various correlation length exponents : we measure the typical correlation exponent $\nu_{typ} \simeq 0.64$ in the disordered phase (this value is very close to the correlation exponent $\nu^Q_{pure}(d=2) \simeq 0.63$ of the {\it pure} two-dimensional quantum Ising Model), and the typical exponent $\nu_h \simeq 1$ within the ordered phase. These values satisfy the relations between critical exponents imposed by the expected finite-size scaling properties at Infinite Disorder critical points. Within the disordered phase, we also measure the fluctuation exponent $\omega \simeq 0.35$ which is compatible with the Directed Polymer exponent $\omega_{DP}(1+1)=\frac{1}{3}$ in $(1+1)$ dimensions.

\end{abstract}

\maketitle

\section{ Introduction }

Strong Disorder Renormalization (see \cite{StrongRGreview} for a review) 
has been first introduced for one-dimensional
quantum spin chains \cite{Ma-Dasgupta,danieltransverse,danielAF},
where exact solutions can be obtained because the renormalized lattice of surviving
degrees of freedom remains one-dimensional.
In dimension $d>1$, the Strong Disorder RG procedure cannot be solved analytically, 
because the topology of the lattice changes upon renormalization,
but it has been implemented numerically, in particular for the quantum Ising model
 \cite{fisherreview,motrunich,lin,karevski,lin07,yu,kovacsstrip,
kovacs2d,kovacs3d,kovacsentropy,kovacsreview}.
Nevertheless, the complicated topology that emerges between renormalized degrees of freedom in dimension $d>1$ tends to obscure the physics and slow down the numerics,
 because a large number of very weak bonds are generated
during the RG, that will eventually not be important for the forthcoming RG steps.
Various types of simplifications have been thus proposed, like
the 'maximum rule' \cite{fisherreview,motrunich,lin,karevski,lin07,yu}
possibly supplemented by some very efficient algorithm \cite{kovacs2d,kovacs3d,kovacsentropy,kovacsreview}, 
the introduction of a cut-off within the full sum rules \cite{iyer}
or the planar approximation \cite{lauman}.
Recently we have proposed to follow another strategy : the idea is to allow some changes in the order of decimations 
with respect to the full procedure in order to maintain a simple spatial renormalized structure. 
We have already applied this idea in two ways :
(i) in \cite{us_boxrg}, we have proposed to include strong disorder RG ideas
within the more traditional fixed-length-scale real space RG framework
that preserves the topology upon renormalization, with numerical results for various types of fractal lattices;
(ii) in \cite{us_treerg}, we have proposed for the Cayley tree geometry
some 'Boundary Strong Disorder RG procedure' that preserves the tree structure, so that one can write simple recursions with respect to the number of 
generations.
In both cases, we have checked that in dimension $d=1$, these modified procedures correctly capture all critical exponents except for the 
magnetic exponent $\beta$ which is related to persistence properties of the full RG flow.
In the present paper, we adapt this idea of  'Boundary Strong Disorder RG procedure' to the two-dimensional
case and present the corresponding numerical results, that we compare with the results of standard
Strong Disorder RG \cite{fisherreview,motrunich,lin,karevski,lin07,yu,kovacsstrip,
kovacs2d,kovacs3d,kovacsentropy,kovacsreview}
and with  quantum Monte-Carlo \cite{pich,rieger}.

The paper is organized as follows.
In section \ref{sec_boundaryrg}, we define the Boundary Strong Disorder RG procedure 
for the two-dimensional square lattice. 
In the following sections, we discuss the numerical results obtained by this procedure.
In the disordered phase (section \ref{sec_disorder}), we measure the typical correlation exponent $\nu_{typ}$,
the fluctuation exponent $\omega$ and the essential singularity exponent $\kappa$.
In the ordered phase (section \ref{sec_order}), we measure the  typical correlation exponent $\nu_{h}$.
In the critical region (section \ref{sec_criti}), we find that the location of the critical point,
the activated exponent $\psi$ and the finite-size correlation exponent $\nu_{FS}$
are compatible with the values obtained previously by standard Strong Disorder RG
\cite{fisherreview,motrunich,lin,karevski,lin07,yu,kovacsstrip,kovacs2d,kovacs3d,kovacsentropy,kovacsreview}.
Our conclusions are summarized in section \ref{sec_conclusion}.

\section{ Boundary Strong Disorder RG procedure in $d=2$ }

\label{sec_boundaryrg}

As recalled in Appendix \ref{app_full}, the Strong Disorder Renormalization
for the quantum Ising model is an energy-based RG, 
where the strongest ferromagnetic bond or the strongest transverse field
is iteratively eliminated. In this section, we introduced a modified procedure,
called Boundary Strong Disorder RG, that preserves a simple spatial structure.

\subsection{ Initial model  }

In this paper, we consider the quantum Ising model defined in terms of Pauli matrices
\begin{eqnarray}
{\cal H} =  -  \sum_{<i,j>} J_{i,j}  \sigma^z_i \sigma^z_j - \sum_i h_i \sigma^x_i
\label{hamilton}
\end{eqnarray}
on the square lattice in dimension $d=2$
where the initial nearest-neighbor couplings $J^{ini}_{i,j} $ are independent random variables drawn with the box distribution on the unit interval $[0,1]$
\begin{eqnarray}
\pi_J(J^{ini}_{ij}) = \theta(0 \leq J^{ini}_{ij} \leq 1)
\label{jdes}
\end{eqnarray}
and
where the initial transverse fields $h^{ini}_i>0$ are independent random variables drawn with 
the box distribution on the interval $[0,h]$
\begin{eqnarray}
\pi_h(h^{ini}_i) = \frac{1}{h} \theta(0 \leq h^{ini}_i \leq h)
\label{hdes}
\end{eqnarray}
so that the parameter $h$ is the control parameter of the quantum phase transition
as in Refs  \cite{lin,karevski,lin07,yu,kovacsstrip,
kovacs2d,kovacs3d,kovacsentropy,kovacsreview,pich,rieger}.

For the numerical results, we consider more precisely a square lattice containing $(2L-1)^2$ spins
of coordinates $(x=1,2,..,2L-1;y=1,2,..,2L-1$. 
Each spin has its random initial transverse field drawn with the distribution of Eq. \ref{hdes}
and is connected to four neighbors via random ferromagnetic couplings drawn
 with the distribution of Eq. \ref{jdes}. All exterior sites situated along the boundaries at $x=0,x=2L$
or $y=0,y=2L$ are identified to a single formal 'external spin' to keep track of the coupling to the boundary
of the finite sample.

\begin{figure}[htbp]
 \includegraphics[height=10cm]{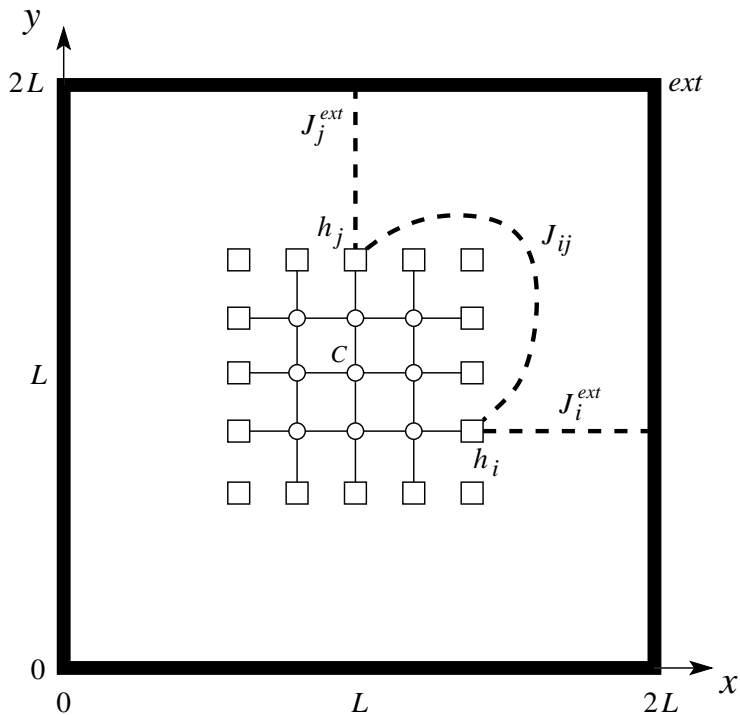}
\caption{ Illustration of the spatial 
renormalized structure during the 'Boundary Strong Disorder RG' in $d=2$ :
any site $k$ belonging to the interior of the corona (denoted here by a circle) has still its initial transverse field $h^{ini}(k)$ and its initial coupling $J^{ini}=J$ to its neighbors;
any site $i$ of the 'corona' (denoted here by a square) has a renormalized transverse field $h_i$,
a renormalized coupling $J^{ext}_i$ to the formal external spin, and possibly a renormalized coupling $J_{ij}$ to any other spin $j$ of the corona; finally the sites outside the corona have been already eliminated, and these eliminations
are responsible of the renormalized variables characterizing the corona sites.
  }
\label{figrg2d}
\end{figure}

\subsection{ Boundary Strong Disorder RG spatial structure  }

We wish to eliminate sites in a simple deterministic order, starting from the boundary :
we will first eliminate sites that are at distance $1$ from the boundary having
coordinates $(x=1)$ or $(x=2L-1)$ or $(y=1)$ or $(y=2L-1)$; then sites that are at distance $2$ 
from the boundary having coordinates
$(x=2)$ or $(x=2L-2)$ or $(y=2)$ or $(y=2L-2)$; and so on, up to sites that are at distance $(L-1)$ 
from the boundary having $(x=L-1)$ or $(x=L+1)$ or $(y=L-1)$ or $(y=L+1)$.
At a given stage of the RG, we have a renormalized spatial structure 
containing a 'corona of renormalized boundary sites' (the number of sites in the corona
scales as the surface $(L-l)^{d-1}$ when the corona is at distance $l$ from the boundary) and the 'interior of the corona' 
(the number of sites in the corona scales as the volume $(L-l)^{d}$ when the corona is at distance $l$ from the boundary).
We have the following properties (see Fig \ref{figrg2d})

- the 'interior of the corona' contains sites that have not yet been modified with respect to the initial model, i.e. the sites are characterized by their initial random fields $h^{ini}(i)$, and are connected to their initial neighbors by their initial ferromagnetic coupling $J^{ini}_{ij}$.

- the 'corona' contains renormalized boundary sites $(i)$ that
 have renormalized transverse fields $h(i)$ and that
 are connected to the formal external spin via some renormalized coupling $J^{ext}(i)$.
These corona sites are connected to 'interior spins' via their initial ferromagnetic coupling $J^{ini}_{ij}$. 
Finally, there may exist renormalized ferromagnetic couplings $J_{ij}$ 
between any two pair $(i,j)$ of sites belonging the corona.

- the sites outside the 'corona' have been already eliminated, and these eliminations
are responsible of the renormalized variables characterizing the corona sites.

\subsection{ Boundary Strong Disorder RG rules for a corona site  }

To eliminate a given site $i$ of the corona, we determine the maximum between
its renormalized transverse field $h(i)$ and its renormalized 
ferromagnetic couplings  $J_{ij}$ with the other sites of the corona or interior sites
\begin{eqnarray}
\Omega_i= {\rm max } \left[h_i,J_{ij} \right]
\label{omegai}
\end{eqnarray}
(Note that the renormalized external coupling $J^{ext}(i)$ is excluded, since it is just a 'passive' variable used to measure the effective coupling to the initial boundary)

Then we apply the Strong Disorder RG rules as follows (see Appendix \ref{app_full}) :

i) If $\Omega_i= h_{i}$, then the site $i$ is decimated, and
 all couples $(j,k)$ of neighbors of $i$ are now linked via
the renormalized ferromagnetic coupling
\begin{eqnarray}
J_{jk}^{new} = J_{jk} + \frac{ J_{ji} J_{ik} }{h_{i}}
\label{jjknewi}
\end{eqnarray}
Accordingly, the external couplings of all neighbors $j$ of $i$ are renormalized according to
\begin{eqnarray}
J_{j}^{ext,new} = J_{j}^{ext} + \frac{ J_{ji} J^{ext}_{i} }{h_{i}}
\label{jjknewext}
\end{eqnarray}

ii) If $\Omega_i= J_{ij}$, then the site $i$ is merged with the site $j$.
The new renormalized site $j$ has a reduced renormalized transverse field
\begin{eqnarray}
h_{j}^{new}= \frac{ h_i h_j}{ J_{ij}  }
\label{hinewi}
\end{eqnarray}
This renormalized site $j$ is connected to other sites $k$ via the renormalized couplings
\begin{eqnarray}
J_{jk}^{new} =  J_{jk}+J_{ik}
\label{couplingjknew}
\end{eqnarray}
In particular, 
the external coupling of the renormalized site $j$ becomes
\begin{eqnarray}
J_{j}^{ext,new} = J_{j}^{ext} +  J^{ext}_{i} 
\label{jjknewexti}
\end{eqnarray}

Note that when the site $i$ is eliminated, all interior sites 
that were connected to the site $i$
become sites of the new corona.

In the final state of the RG procedure, only the center site of coordinates $(x_c=L,y_c=L)$ remains. The observables under interest are its final renormalized transverse-field $h_L$ and its ferromagnetic coupling $J_L^{ext}$ to the formal 'external spin', i.e. to the initial boundary of the square sample.
To simply the notations, $ J_L^{ext}$ will be denoted simply by $J_L $ from now on.
In the following, we will concentrate on the typical values $h_L^{typ}$ and $J_L^{typ}$
defined by 
\begin{eqnarray}
\ln h_L^{typ} && \equiv \overline{\ln h_L} \nonumber \\
\ln J_L^{typ} && \equiv \overline{\ln J_L} 
\label{deftyp}
\end{eqnarray}
and on the widths of the distribution of $\ln h_L$ and $\ln J_L$ defined by
\begin{eqnarray}
\Delta_{\ln h_L}  && \equiv  \left( \overline{ (\ln h_L)^2 } - (\overline{ \ln h_L })^2 \right)^{1/2}  
\nonumber \\
\Delta_{\ln J_L}  && \equiv  \left( \overline{ (\ln J_L)^2 } - (\overline{ \ln J_L })^2 \right)^{1/2}  
\label{defwidth}
\end{eqnarray}
where the overbar denotes an average over the disordered samples.

\subsection{ Numerical details  }

We have followed numerically the Boundary RG rules for square samples containing $N_L=(2L-1)^2$ spins,
for various sizes $L \leq 100$ corresponding to $N_L \leq 39601$.
For a given size $L$, the number $n_s(L)$ of independent disordered samples we have been able to study
depends on the value $h$ of the initial disorder distribution of Eq. \ref{hdes}.
So let us give some typical values we have used in the critical region 
(we were able to study more samples in the ordered phase $h<h_c$ and less samples in the disordered phase)
\begin{eqnarray}
 L && = 6, 10, 20, 40, 60, 80, 100 \nonumber \\
n_s(L) && = 10^8, 2.10^7, 3.10^5, 2.10^4 , 2.10^3, 5.10^2, 2.10^2
\label{nume}
\end{eqnarray}
Our various data shown below are compatible
 with a critical point located around the value
(see the definition of the control parameter $h$ in the initial disorder distribution of Eq. \ref{hdes})
\begin{eqnarray}
      h_c \simeq 5.15
\label{hcregion}
\end{eqnarray}
which is sligthly lower but close
 to the values found previously using the standard Strong Disorder RG rules with the maximum rule,
namely $h_c \simeq 5.3 $ \cite{lin} and $h_c \simeq 5.35 $ \cite{yu,kovacsstrip,kovacs2d,kovacsentropy,kovacsreview}. 
Let us first discuss the properties of the two phases, before we turn to the critical region.

\section{ Analysis of the disordered phase }

\label{sec_disorder}

\subsection{ RG flow of the renormalized external coupling $J_L$ in the disordered phase}

\begin{figure}[htbp]
 \includegraphics[height=6cm]{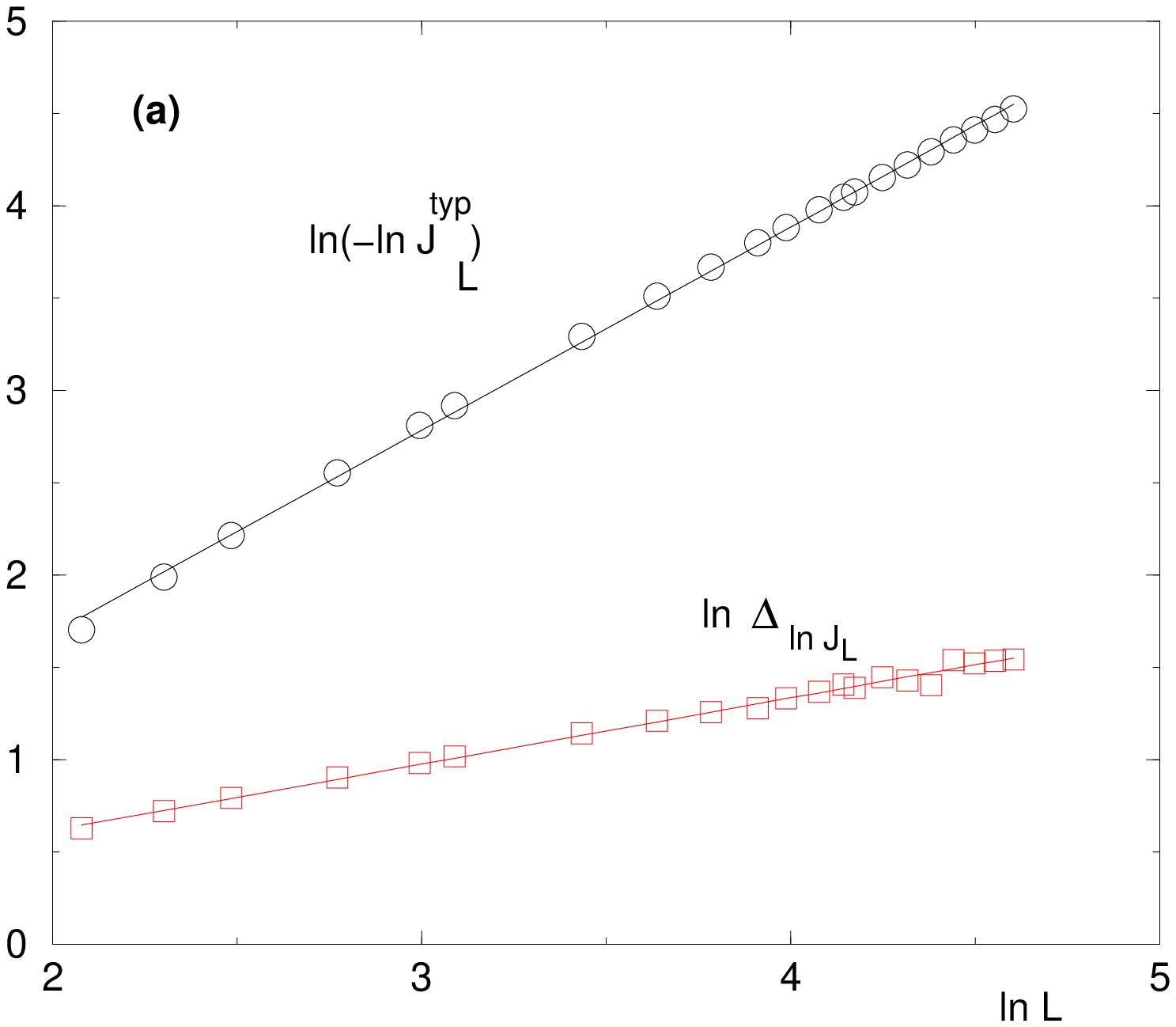}
\hspace{1cm}
 \includegraphics[height=6cm]{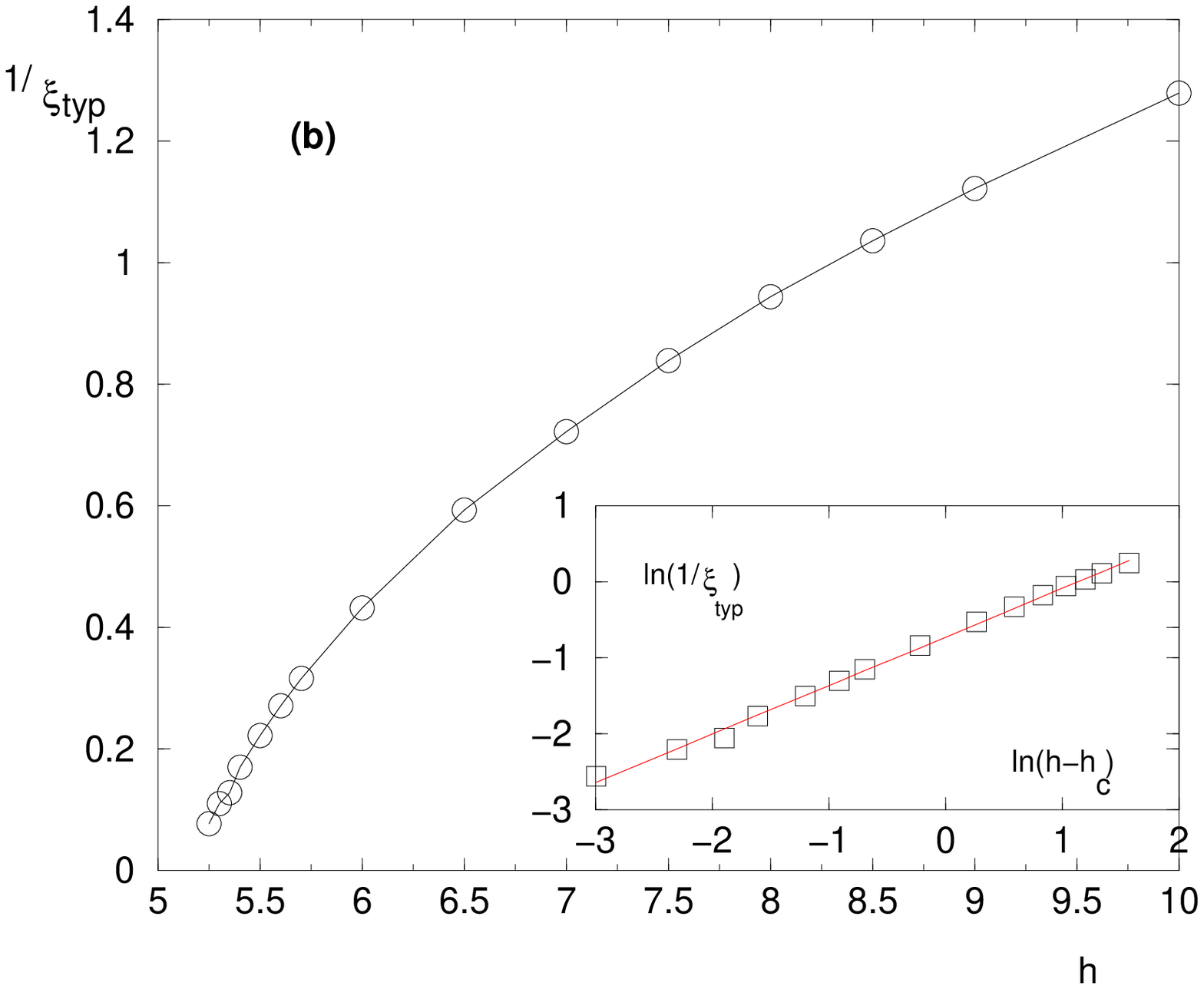}
\caption{ 
(a)  Statistics of the logarithm of the external renormalized coupling $(\ln J_L)$ in the disordered phase (here $h=8$) : 
the RG flows of the typical value $\ln J_L^{typ}$ 
and of the width $\Delta_{\ln J_L} $ in a log-log plot display 
 slopes $1$ and $\omega \simeq 0.35$ respectively (Eq. \ref{JLdisorderomega})
(b) Inverse of the typical correlation length $\xi_{typ}$ of Eq. \ref{JLdisordertyp}
as a function of the control parameter $h>h_c$: the correlation length exponent $\nu_{typ}$ of Eq. \ref{nutypdes}
is of order $\nu_{typ} \simeq 0.64$ (see the log-log plot in inset).
  }
\label{figjdes}
\end{figure}

In the disordered phase, 
the renormalized external coupling $J_{L}$ is expected to 
present the following scaling
\begin{eqnarray}
\ln J_L = - \frac{L}{\xi_{typ}} + L^{\omega} A(h) u
\label{JLdisorder}
\end{eqnarray}
The first non-random term describing the exponential decay with the size $L$ ( see Fig. \ref{figjdes} (a) )
 defines the typical correlation length $\xi_{typ}$
\begin{eqnarray}
\ln J_L^{typ} \equiv \overline{\ln J_L} \oppropto_{L \to +\infty} - \frac{L}{\xi_{typ}} 
\label{JLdisordertyp}
\end{eqnarray}
On Fig. \ref{figjdes} (b), we show how $1/\xi_{typ}$ varies as a function of the control parameter $h$
of the transition : our data are compatible with the power-law divergence 
(see the log-log plot in the inset of  Fig. \ref{figjdes} (b))
\begin{eqnarray}
\xi_{typ} \propto (h-h_c)^{-\nu_{typ}}
\label{nutypdes}
\end{eqnarray}
with a typical correlation exponent of order 
\begin{eqnarray}
\nu_{typ} \simeq 0.64
\label{nutypnume}
\end{eqnarray}
To the best of our knowledge, this is the first numerical measure of this typical exponent $\nu_{typ}$
within the disordered phase,
since previous studies have concentrated on the critical region where finite-size effects are governed
by another correlation length exponent $\nu_{FS}$ (see section \ref{sec_criti}).
We note that the value found here for the typical exponent of Eq. \ref{nutypnume}
turns out to be very close to the correlation exponent $\nu^Q_{pure}(d=2) \simeq 0.63$ of the 
{\it  pure two-dimensional quantum Ising Model }. (The latter is known to coincide with the correlation exponent 
$\nu^{class}_{pure}(d+1=3) \simeq 0.63$
 of the {\it pure  three-dimensional classical Ising Model } as a consequence of the quantum-classical correspondence
\cite{sachdev}). Since in dimension $d=1$, the typical exponent $\nu_{typ}(d=1)=1$ also coincides with the 
correlation exponent $\nu^Q_{pure}(d=1) =1$ of the 
{\it  pure one-dimensional quantum Ising Model } (and equivalently with the exponent
$\nu^{class}_{pure}(d+1=2)= 1$ of the {\it pure  two-dimensional classical Ising Model }), 
it would be interesting to determine whether these coincidences continue in higher dimensions $d \geq 3$,
i.e. whether the typical exponent takes the simple value $\nu_{typ}(d \geq 3)=\nu^Q_{pure}(d \geq 3)=\nu^{class}_{pure}(d+1\geq 4)=\frac{1}{2} $ ?

The second term in Eq \ref{JLdisorder} contains an $O(1)$ random variable $u$,
which is expected to be subleading with respect to the first term,
i.e. the width $\Delta_{\ln J_L}$ of the distribution of $\ln J_L$
is of order $L^{\omega}$ with some fluctuation exponent $\omega<1$
\begin{eqnarray}
\Delta_{\ln J_L}  \equiv  \left( \overline{ (\ln J_L)^2 } - (\overline{ \ln J_L })^2 \right)^{1/2} \oppropto_{L \to +\infty} L^{\omega} 
\label{defdeltaLomega}
\end{eqnarray}
We have argued in \cite{us_transverseDP} that
this exponent $\omega$ should coincide with the droplet exponent $\omega_{DP}(D=d-1)$
 of the Directed Polymer with $D=(d-1)$ transverse directions.
For our present case in $d=2$, the droplet exponent of the Directed Polymer is 
exactly known to be $\omega_{DP}(D=1)=\frac{1}{3}$
\begin{eqnarray}
 \omega=\omega_{DP}(D=1)=\frac{1}{3}
\label{JLdisorderomega}
\end{eqnarray}
in agreement with our numerical results shown on Fig. \ref{figjdes} (a).
Again, to the best of our knowledge, this fluctuation exponent $\omega$
within the disordered phase had not been measured yet,
since previous studies have concentrated on the critical region.

\subsection{ RG flow of the renormalized transverse field $h_L$ in the disordered phase}

\begin{figure}[htbp]
 \includegraphics[height=6cm]{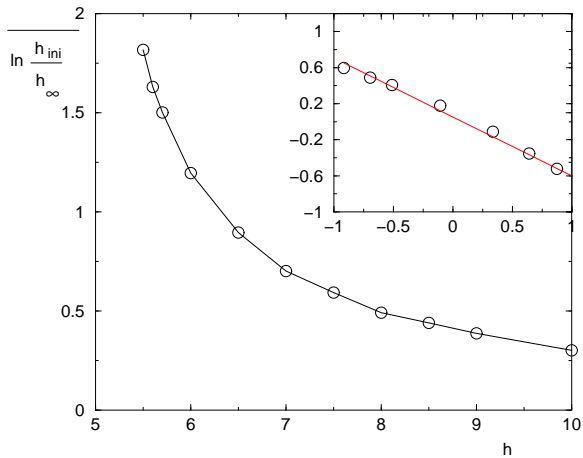}
\hspace{1cm}
\caption{ 
  Statistics of the asymptotic finite renormalized transverse field $h_{L \to \infty}$
in the disordered phase as a function of the control parameter $h>h_c$ :
the critical exponent $\kappa$ of Eq. \ref{defkappa} is of order $\kappa \simeq 0.65$ 
(see the log-log plot in inset).
  }
\label{fighdes}
\end{figure}

In the disordered phase, the renormalized transverse field $h_L$ is expected
to remain a finite random variable as $L \to +\infty$.
In particular, the typical value remains finite
\begin{eqnarray}
\ln h_{L}^{typ} \equiv \overline{\ln h_{L}} \oppropto_{L \to +\infty}  \overline{\ln h_{\infty}}
\label{lnhlfinite}
\end{eqnarray}
To analyze more clearly the statistics of the finite renormalization with respect
to the initial random fields $h_{ini}$ drawn with the box distribution of Eq. \ref{hdes}
corresponding to
\begin{eqnarray}
\overline{ \ln h_{ini}} = \int_0^h  \frac{dh_{ini}}{h} \ln h_{ini} = \ln  h -1
\label{lnhdes}
\end{eqnarray}
we show on Fig. \ref{fighdes}  the difference $(\overline{\ln h_{\infty}} - \overline{ \ln h_{ini}})$
as a function of $h>h_c$. Our data are compatible with the expected essential singularity
\begin{eqnarray}
\overline{ \ln \frac{ h_{\infty}} {  h_{ini}}  }  \propto   - (h-h_c)^{-\kappa} 
\label{defkappa}
\end{eqnarray}
with an exponent of order
\begin{eqnarray}
\kappa \simeq 0.65
\label{numekappa}
\end{eqnarray}

Since the dynamical exponent $z$ is expected to have the same singularity as the averaged value of Eq. \ref{defkappa},
we may compare our estimate of $\kappa$ with the measure given after Eq (61) of Ref. \cite{kovacs2d} based on standard RSRG procedure: $ z \propto (h-h_c)^{0.60(6)}$.

\section{ Analysis of the ordered phase }

\label{sec_order}

\subsection{ RG flow of the renormalized transverse field $h_L$  in the ordered phase}

\begin{figure}[htbp]
 \includegraphics[height=6cm]{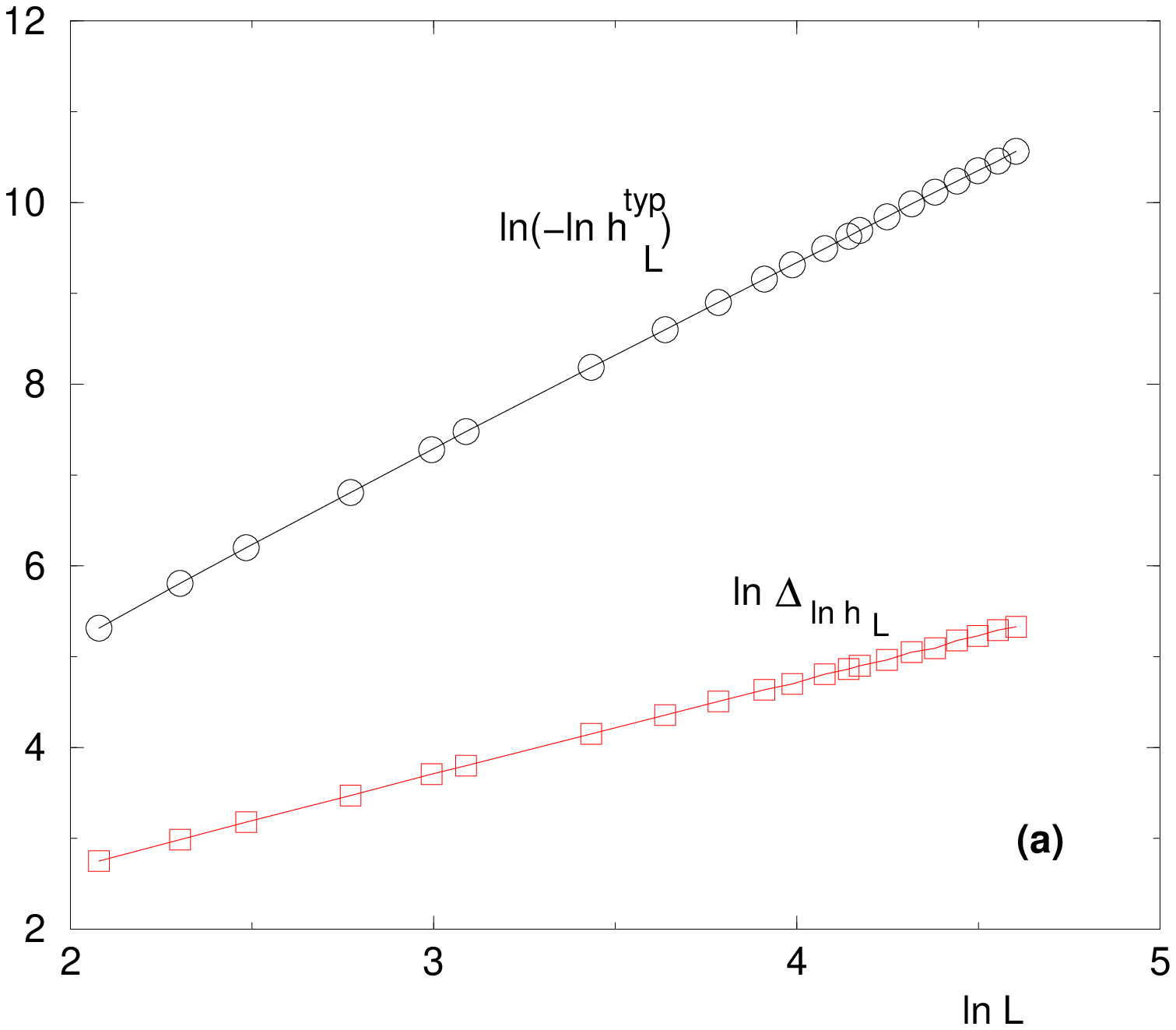}
\hspace{1cm}
 \includegraphics[height=6cm]{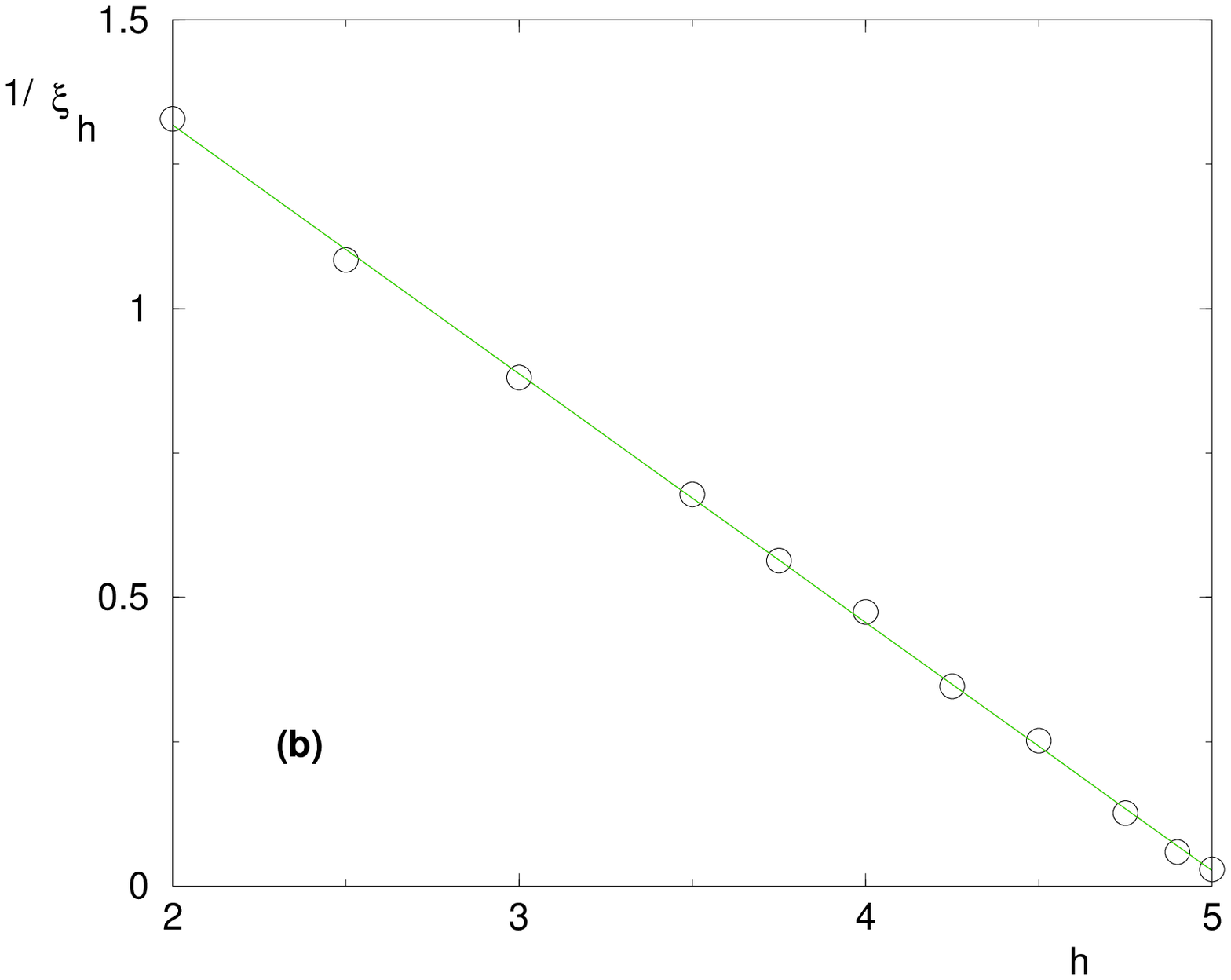}
\caption{ 
(a)  Statistics of the logarithm of the renormalized transverse field $(\ln h_L)$ in the ordered phase (here $h=1$) : 
the RG flows of the typical value $(\ln h_L^{typ})$
and of the width $\Delta_{\ln h_L} $ in a log-log plot display 
 slope $d=2$ and slope $d/2=1$ respectively.
(b) Inverse of the correlation length $\xi_h$ of Eq. \ref{hLorder}
as a function of the control parameter $h<h_c$ : the correlation length exponent $\nu_h$ of Eq. \ref{nutypdes}
is of order $\nu_h \simeq 1$.  }
\label{fighorder}
\end{figure}

In the ordered phase, 
the logarithm of the renormalized transverse field is expected to behave extensively in the volume $L^d$
\begin{eqnarray}
\ln h_{L}^{typ} \equiv \overline{\ln h_{L}}
 \oppropto_{L \to \infty}   - \left( \frac{L}{\xi_h} \right)^d 
\label{hLorder}
\end{eqnarray}
with $d=2$, in agreement with our data shown on Fig. \ref{fighorder} (a).
The length scale $\xi_h$ represents the characteristic size of finite disordered clusters
within this ordered phase. It is expected to diverge as a power-law near the transition
\begin{eqnarray}
\xi_{h} \propto (h_c-h)^{-\nu_h}
\label{nuhorder}
\end{eqnarray}
The corresponding correlation exponent  $\nu_h$
plays in the ordered phase a role similar to $\nu_{typ}$ in the disordered phase (Eq. \ref{nutypdes}).
Our data are compatible with a value of order (see Fig. \ref{fighorder} (b))
\begin{eqnarray}
\nu_h \simeq 1
\label{nuhnume}
\end{eqnarray}
Again, to the best of our knowledge, this is the first numerical measure
 of the typical exponent $\nu_{h}$ within the ordered phase,
since previous studies have concentrated on the critical region where finite-size effects are governed
by another correlation length exponent $\nu_{FS}$ (see section \ref{sec_criti}).
We note that this exponent takes also the same value $\nu_h(d=1)=1$ in $d=1$,
but we are not aware of any argument in favor of this simple constant value as $d$ varies.

As shown on Fig. \ref{fighorder} (a), the width of the distribution of the logarithm of the renormalized transverse field
grows linearly in $L$
\begin{eqnarray}
\Delta_{\ln h_L}  \equiv  \left( \overline{ (\ln h_L)^2 } - (\overline{ \ln h_L })^2 \right)^{1/2} \oppropto_{L \to +\infty} L
\label{defdeltahL}
\end{eqnarray}

\subsection{ RG flow of the renormalized external coupling $J_L$  in the ordered phase}

\begin{figure}[htbp]
 \includegraphics[height=6cm]{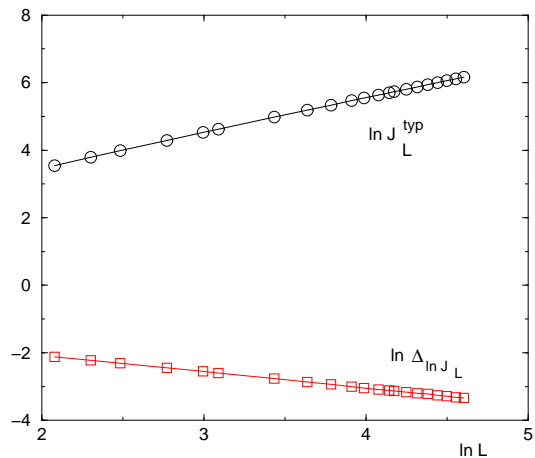}
\caption{ 
  Statistics of the renormalized external coupling $J_L$ in the ordered phase (here $h=1$) : 
the RG flows of the typical value $(\ln J_L^{typ})$ of Eq. \ref{jlclassical}
and of the log of the width $\ln (\Delta_{\ln J_L}) $
of Eq. \ref{defdeltahL} display 
 slope $1$ and slope $(-1/2)$ respectively. 
  }
\label{figjorder}
\end{figure}

In dimension $d=1$ where there is no underlying classical ferromagnetic transition, the typical renormalized coupling remains finite in the ordered phase, and presents the same essential singularity as in Eq. \ref{defkappa}.
However in dimension $d>1$ where there exists
 an underlying classical ferromagnetic transition,
the renormalized couplings $J_L$ is expected to grow at large $L$
with the scaling of the {\it classical } random ferromagnetic
 model (see Fig. \ref{figjorder} )
\begin{eqnarray}
\ln J_L^{typ} \equiv \overline{ \ln J_L} \oppropto_{L \to +\infty}   \ln \left( \sigma(h) L^{d_s} \right)
\label{jlclassical}
\end{eqnarray}
where $d_s=d-1=1$ represents the interface dimension, and where 
$\sigma$ represents the surface tension.
We are able to measure the asymptotic behavior of Eq. \ref{jlclassical} only
sufficiently far  $h \leq 4.25$ from the critical point $h_c \simeq 5.15$,
so that we cannot measure the critical behavior of the surface tension $\sigma(h)$.

As shown on Fig. \ref{figjorder}, the width $ \Delta_{\ln J_L}$ of the distribution of
the logarithm of the external renormalized coupling decays as 
\begin{eqnarray}
 \Delta_{\ln J_L} \oppropto_{L \to +\infty} L^{-0.5}
\label{widthjlorder}
\end{eqnarray}

\section{ Analysis of the critical region }

\label{sec_criti}

\subsection{ RG flow of the  renormalized transverse field at criticality  }

\begin{figure}[htbp]
 \includegraphics[height=6cm]{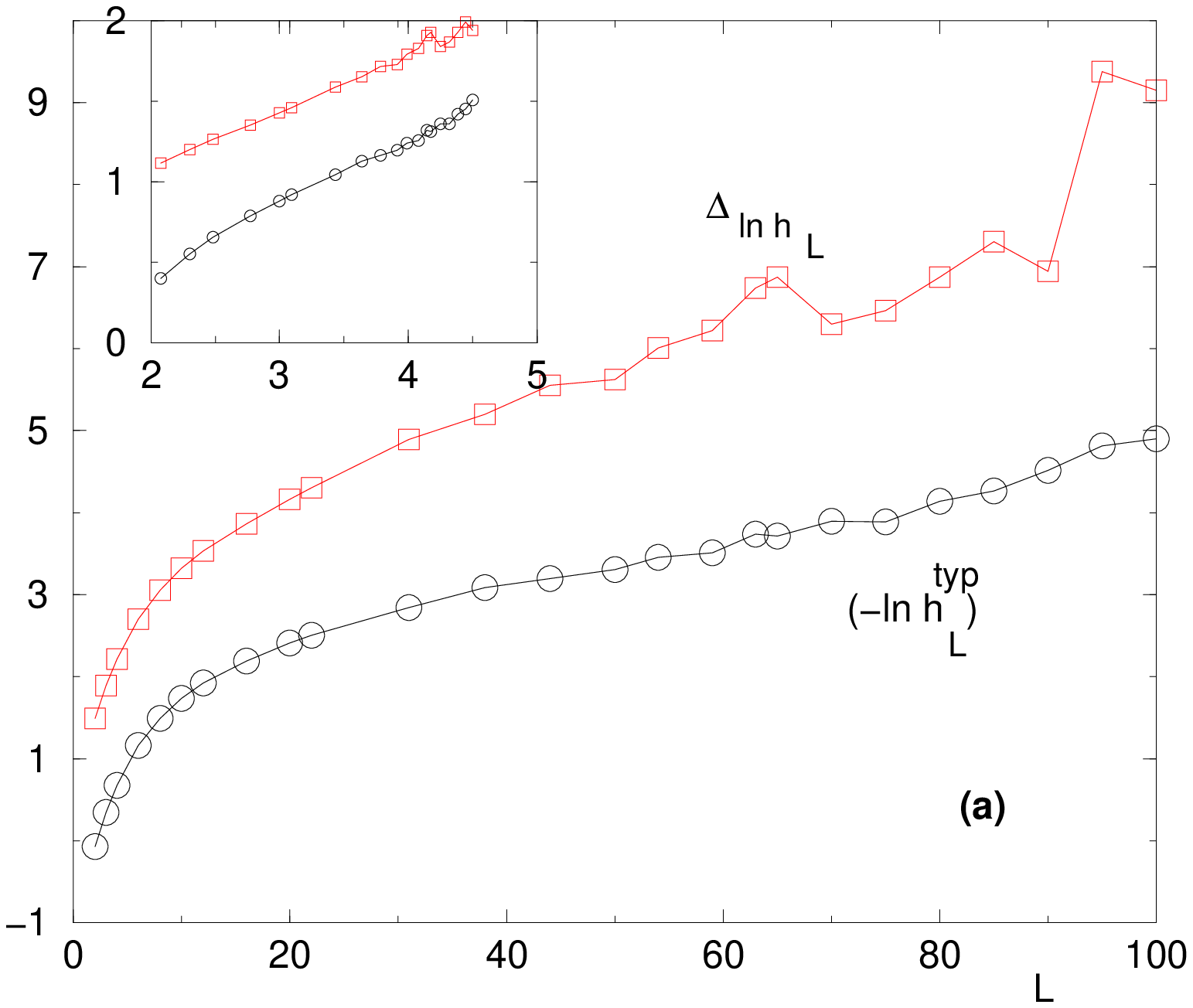}
\hspace{1cm}
 \includegraphics[height=6cm]{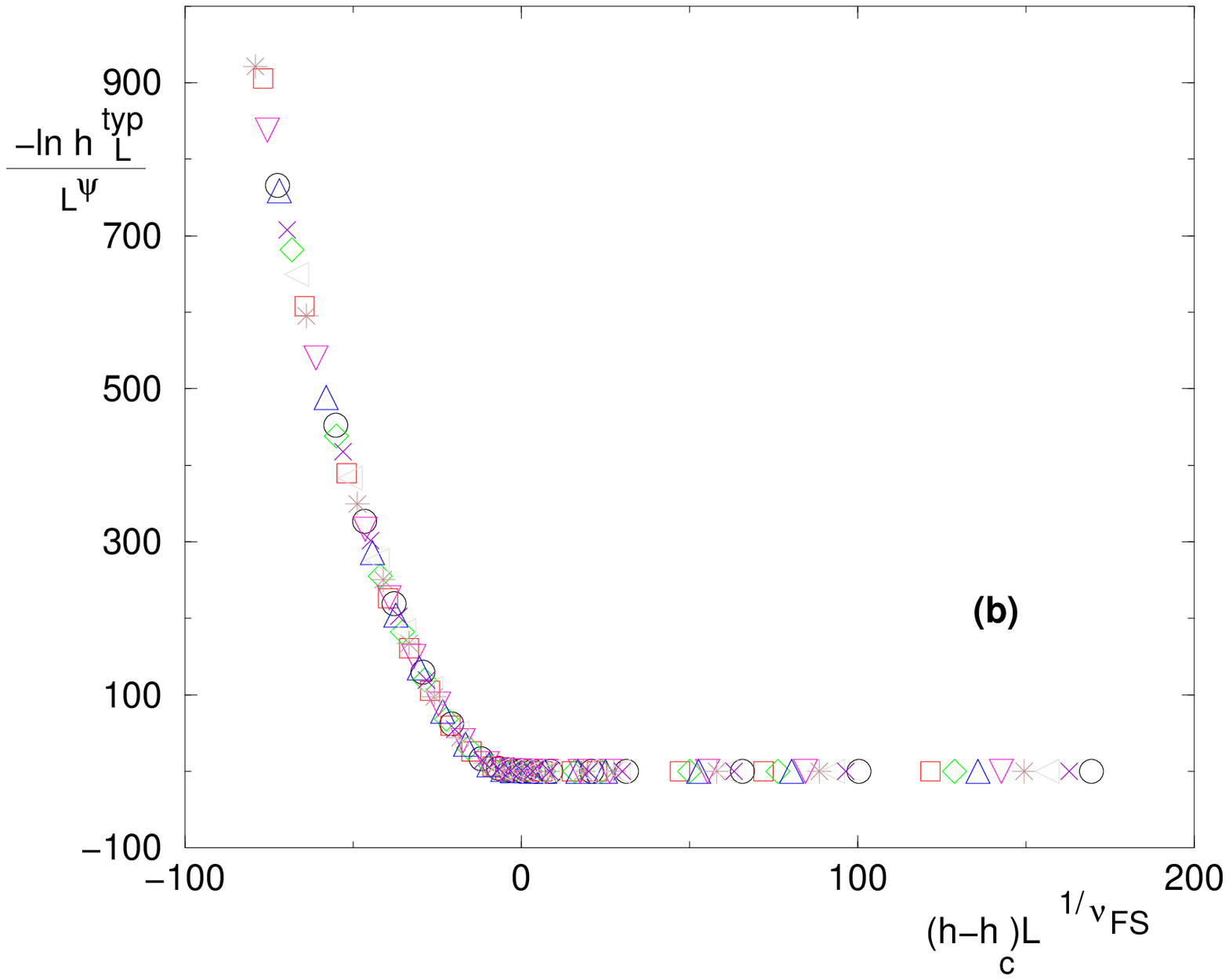}
\caption{ 
(a)  Statistics of the renormalized transverse field $h_L$ at criticality $h_c \simeq 5.15$ :
$(-\ln h_L^{typ})$ and $\Delta_{\ln h_L} $ as a function of $L$ (Inset: log-log plot of the same data)
(b) Test of the finite-size scaling form of Eq. \ref{fsslnhl} : the data collapse
obtained with $\psi \simeq 0.5$ and $\nu_{FS} \simeq 1.3$ is satisfactory  }
\label{fighcriti}
\end{figure}

At the Infinite Disorder critical point,
 the renormalized transverse field $h_L$ is expected to display
an activated scaling in $L$ with some exponent $\psi$
\begin{eqnarray}
\ln h_L && \propto -  L^{\psi} v_c
\label{JLhLcriti}
\end{eqnarray}
where $v_c$ is an $O(1)$ random variable.
We show on Fig. \ref{fighcriti} (a) 
our data concerning the RG flows of the typical value $(\overline {\ln h_L }) $ and the width $\Delta_{\ln h_L}$
of the distribution of the logarithm of the renormalized transverse field at criticality $h=h_c\simeq 5.15$.
Our data are consistent with the same scaling of both,
with an exponent in the region
\begin{eqnarray}
0.4 \leq \psi \leq  0.5
\label{psicriti}
\end{eqnarray}
in agreement with previous estimates based on standard Strong Disorder RG
\cite{fisherreview,motrunich,lin,karevski,lin07,yu,kovacsstrip,
kovacs2d,kovacs3d,kovacsentropy,kovacsreview}
or on  quantum Monte-Carlo \cite{pich,rieger}.
The numerical estimate of Eq. \ref{psicriti} from the scaling at criticality
is not precise as a consequence of the uncertainty
of the exact location of the critical point (Eq. \ref{hcregion}),
and of the curvature of the data in log-log plots.
It is thus interesting to discuss the finite-size scaling in the critical region
to relate $\psi$ to other critical exponents of the ordered and disordered phases measured in previous sections.

In the critical region around this Infinite Disorder fixed point, one expects
the following finite-size scaling form for the typical values \cite{motrunich}
\begin{eqnarray}
\ln h_L^{typ} \equiv \overline{ \ln h_L }  = -  L^{\psi} F_h\left( L^{1/\nu_{FS}} \vert h-h_c \vert \right)
\label{fsslnhl}
\end{eqnarray}
where $\nu_{FS}$ is the correlation length exponent that govern all finite-size effects in
the critical region. (The exponent $\nu_{FS}$ is expected \cite{danieltransverse,motrunich}
to correspond to the exponent $\nu_{av}$ of the {\it averaged } two-point correlation function $\nu_{FS}=\nu_{av}$).

The compatibility of the finite-size scaling form of Eq. \ref{fsslnhl}
with the essential singularity of Eq. \ref{defkappa} concerning the disordered phase yields the relation
\begin{eqnarray}
\kappa = \psi \nu_{FS}
\label{kappafss}
\end{eqnarray}
whereas the compatibility with the behavior of Eq. \ref{hLorder} concerning the ordered phase yields the relation
\begin{eqnarray}
\nu_{h} = \left( 1- \frac{\psi}{d}\right) \nu_{FS}
\label{nuh}
\end{eqnarray}
Eliminating $\nu_{FS}$, we may thus obtain a numerical estimate of $\psi$
from our previous measures of $\kappa \simeq 0.65$ (Eq \ref{numekappa})
and $\nu_h \simeq 1 $ (Eq \ref{nuhnume})
\begin{eqnarray}
\psi = \frac{\kappa}{\nu_h+ \frac{\kappa}{d}} \simeq 0.49
\label{psifromkappanuh}
\end{eqnarray}
and the corresponding finite-size exponent then reads
\begin{eqnarray}
\nu_{FS} = \nu_h+ \frac{\kappa}{d} \simeq 1.32
\label{nufsfromkappanuh}
\end{eqnarray}
These two values are close to the values $\psi \simeq 0.48$ and  $\nu \simeq 1.25$ obtained by standard Strong Disorder RG 
\cite{kovacsstrip,kovacs2d,kovacsentropy,kovacsreview}.
To test Eq. \ref{fsslnhl}, we show on Fig \ref{fighcriti} (b) the satisfactory
data collapse
obtained with $\psi \simeq 0.5$ and $\nu_{FS} \simeq 1.3$.

\subsection{ RG flow of the   renormalized external coupling $J_L$ at criticality }

\begin{figure}[htbp]
 \includegraphics[height=6cm]{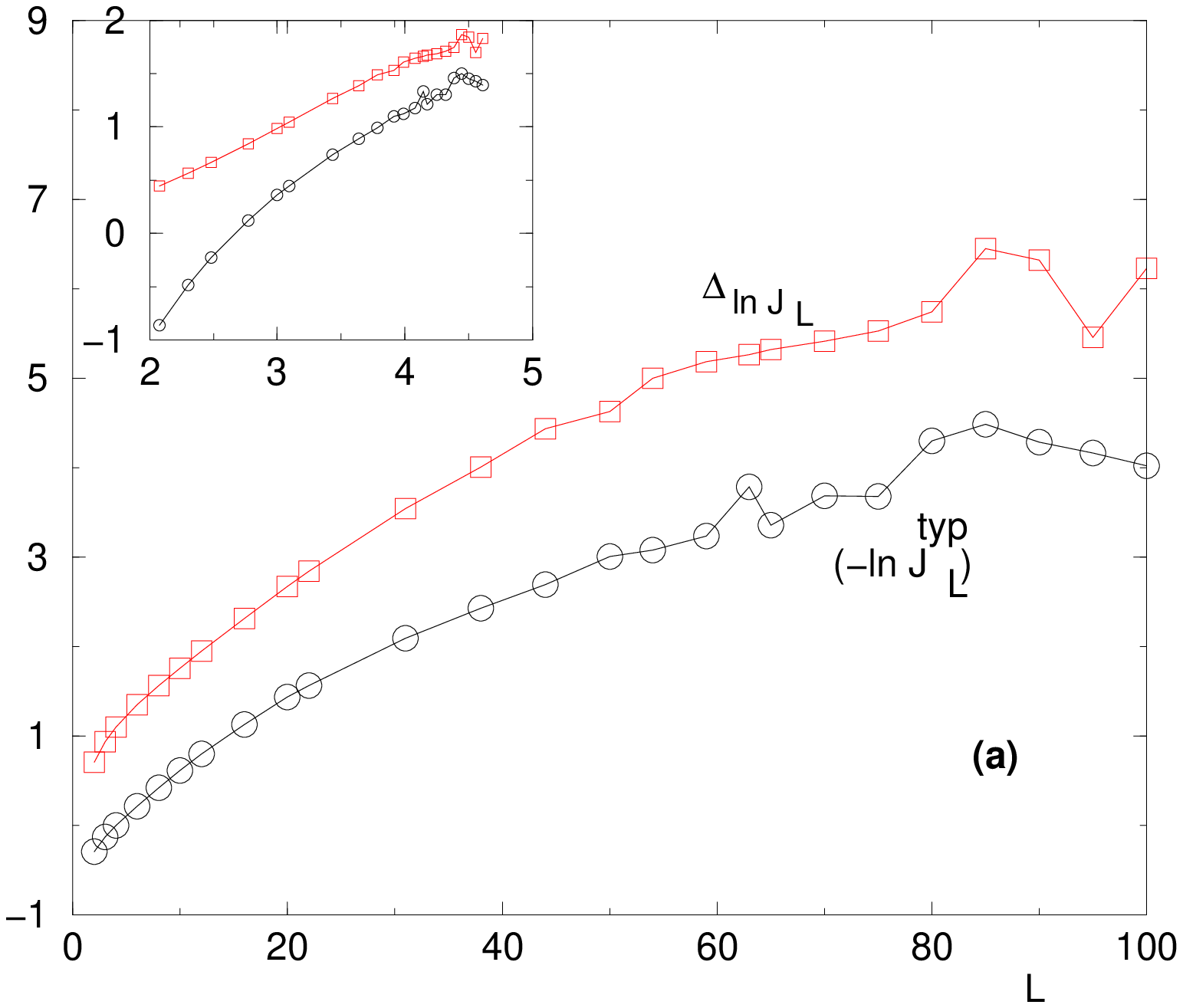}
\hspace{1cm}
 \includegraphics[height=6cm]{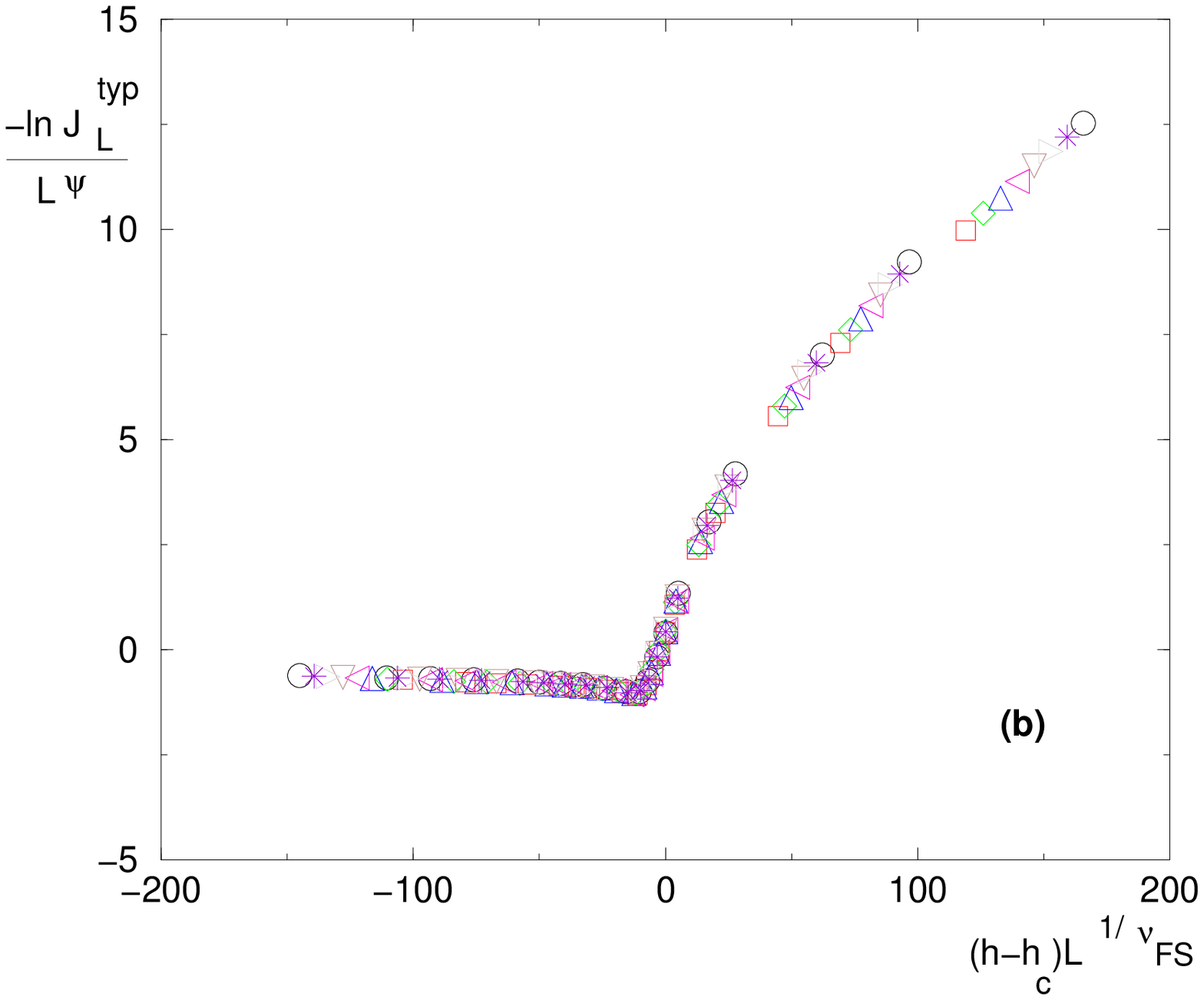}
\caption{ 
(a)   Statistics of the renormalized external coupling $J_L$ 
at criticality $h_c \simeq 5.15$:
$ (- \ln J_L^{typ})$ and $\Delta_{\ln J_L}$ as a function of $ L$
  (Inset: log-log plot of the same data) 
(b)Test of the finite-size scaling form of Eq. \ref{fsslnjl} : the data collapse
obtained with $\psi \simeq 0.5$ and $\nu_{FS} \simeq 1.3$ is satisfactory. }
\label{figjcriti}
\end{figure}

We show on Fig. \ref{figjcriti} (a) 
our data concerning the RG flows of the typical value $\ln J_L^{typ} $ and of the width $\Delta_{\ln J_L}$
of the distribution of the logarithm of the renormalized external coupling at criticality $h=h_c \simeq 5.15$.
Our data are again consistent with the same scaling of both,
with an exponent $\psi$ again in the interval of Eq. \ref{psicriti}.

To estimate a more precise value, it is again interesting to consider the finite-size scaling
properties in the critical region
\cite{motrunich}
\begin{eqnarray}
\ln J_L^{typ} \equiv \overline{ \ln J_L }  = -  L^{\psi} F_J\left( L^{1/\nu_{FS}} \vert h-h_c \vert \right)
\label{fsslnjl}
\end{eqnarray}
  The compatibility the finite-size scaling form with the behavior of Eq. \ref{JLdisordertyp}
concerning the disordered phase implies the following relation between exponents
\begin{eqnarray}
\nu_{typ} = (1-\psi) \nu_{FS}
\label{nutypnufs}
\end{eqnarray}

The comparison with the relation of Eq. \ref{kappafss} yields 
with our previous measures of $\nu_{typ} \simeq  0.64 $
of Eq. \ref{nutypnume} and of  $\kappa \simeq 0.65$ (Eq \ref{numekappa})
\begin{eqnarray}
\psi = \frac{1}{1+\frac{\nu_{typ}}{\kappa} } \simeq 0.5
\label{psifromnutypkappa}
\end{eqnarray}
in agreement with Eq. \ref{psifromkappanuh}.
The corresponding value 
\begin{eqnarray}
\nu_{FS} =\nu_{typ} +\kappa \simeq 1.29
\label{nufsbis}
\end{eqnarray}
 is close to the value of Eq. \ref{nufsfromkappanuh} and to to value 
$\nu \simeq 1.25$ obtained by standard Strong Disorder RG 
\cite{kovacsstrip,kovacs2d,kovacsentropy,kovacsreview}.
To test Eq. \ref{fsslnjl}, we show on Fig \ref{figjcriti} (b) the  satisfactory
data collapse
obtained with $\psi \simeq 0.5$ and $\nu_{FS} \simeq 1.3$.

As a final remark, we have thus found that the activated exponent $\psi \simeq 0.5$ at criticality
is greater than the fluctuation exponent $\omega =1/3$ of the disordered phase (Eq. \ref{JLdisorderomega}),
as already found on fractal lattices in $d>1$ \cite{us_boxrg}, 
whereas in dimension $d=1$,
the two exponents coincide $\psi(d=1)=1/2=\omega(d=1)$.
This means that the amplitude $A(h)$ of Eq. \ref{JLdisorder}
should diverge as $A(h) \propto (h-h_c)^{-(\psi-\omega)\nu_{FS}}$, but we are not able to measure this singularity with our data.

\section{ Conclusion  }

\label{sec_conclusion}

To avoid the complicated topology of surviving clusters induced by standard Strong Disorder RG in dimension $d>1$,
we have introduced a modified procedure called 'Boundary Strong Disorder RG'
 for the Random Transverse Field Ising model in $d=2$. The hope is that, 
as for the one-dimensional case discussed in \cite{us_treerg},
this simpler  'Boundary Strong Disorder RG' that changes the order of decimations
could be able to reproduce correctly all critical exponents except 
 the magnetic exponent $\beta$ which is related to persistence properties of the RG flow.
Note that within the standard RSRG procedure, the effects of 'bad decimations' 
in one-dimension
 have been analysed in detail in Appendix E of Ref. \cite{danieltransverse}, with the conclusion that 
'we recover exactly at a later stage from the errors made earlier'.
This robustness of strong disorder RG rules against mistakes in the order of decimations
has been also found in higher dimensional systems in another context
(see section 3.6 of \cite{us_valley}).
Here we thus hope that this phenomenon still occurs when one imposes even more bad decimations.
However, 
since no exact result exists for RSRG in $d=2$, we cannot really prove that this hope is correct,
but we have presented detailed numerical results obtained by the 'Boundary Strong Disorder RG' 
to compare them with previous results obtained via standard RSRG.

We have found that the location of the critical point,
the activated exponent $\psi \simeq 0.5$ of the Infinite Disorder scaling, 
and the finite-size correlation exponent $\nu_{FS} \simeq 1.3$
are compatible with the values obtained previously by standard Strong Disorder RG
\cite{fisherreview,motrunich,lin,karevski,lin07,yu,kovacsstrip,
kovacs2d,kovacs3d,kovacsentropy,kovacsreview}. 
We thus believe that our modified simplified procedure
captures correctly the critical properties. 
 In addition, we have analyzed in detail the RG flows within the two phases.
Within the disordered phase, we have measured  the typical correlation exponent
$\nu_{typ} \simeq 0.64$, which is very close to the correlation exponent $\nu^Q_{pure}(d=2) \simeq 0.63$ of the {\it pure} two-dimensional quantum Ising Model;
  the fluctuation exponent $\omega \simeq 0.35$ which is compatible with the Directed Polymer exponent $\omega_{DP}(1+1)=\frac{1}{3}$ in $(1+1)$ dimensions, in agreement with the arguments of Ref \cite{us_transverseDP};
 the essential singularity exponent $\kappa \simeq 0.65$.
Within the ordered phase, we have measured the typical exponent $\nu_h \simeq 1$, 
which is close to the value $\nu_h(d=1)=1$ in $d=1$.

The simple values found here for $\nu_{typ}$ and $\nu_h$ in $d=2$, together with the exact solution in $d=1$,
raise the question whether, in higher dimensions $d \geq 3$, the typical exponent $\nu_{typ}$ still
coincides with the correlation exponent of the pure quantum transition 
$\nu_{typ}(d \geq 3) = \nu^Q_{pure}(d \geq 3)=1/2$ ? and whether the typical exponent $\nu_h$ still
keeps the simple value $\nu_h(d \geq 3) \simeq 1$ ?

More generally, we hope that the idea to change the order of decimations
with respect to the standard Strong Disorder RG
in order to simplify the renormalized spatial structure will be useful 
for all types of models controlled by Infinite Disorder scaling.

\appendix

\section{ Reminder on Strong Disorder RG rules on arbitrary lattices }

\label{app_full}

For the Random Transverse Field Ising Model of Eq. \ref{hdes},
we recall that the standard Strong Disorder Renormalization 
 are formulated on arbitrary lattices as follows \cite{fisherreview,motrunich} :

(0) Find the maximal value among all the transverse fields $h_i$
and all the ferromagnetic couplings $J_{jk}$
\begin{eqnarray}
\Omega= {\rm max } \left[h_i,J_{jk} \right]
\label{omega}
\end{eqnarray}

i) If $\Omega= h_{i}$, then the site $i$ is decimated and disappears,
while all couples $(j,k)$ of neighbors of $i$ are now linked via
the renormalized ferromagnetic coupling
\begin{eqnarray}
J_{jk}^{new} = J_{jk} + \frac{ J_{ji} J_{ik} }{h_{i}}
\label{jjknew}
\end{eqnarray}

ii) If $\Omega= J_{ij}$, then the site $j$ is merged with the site $i$.
The new renormalized site $i$ has a reduced renormalized transverse field
\begin{eqnarray}
h_{i}^{new}=  \frac{h_i h_j}{ J_{ij}  }
\label{hinew}
\end{eqnarray}
and is connected to other sites via the renormalized couplings
\begin{eqnarray}
J_{ik}^{new} =  J_{ik}+J_{jk}
\label{jiknew}
\end{eqnarray}

(iii) return to (0).

These standard Strong Disorder RG rules should be compared with the 
modified procedure called 'Boundary  Strong Disorder RG' introduced in section \ref{sec_boundaryrg}.

\end{document}